\documentclass[aps,prc,preprint,tightenlines,
superscriptaddress,nofootinbib,14pt]{revtex4}
\usepackage{graphicx}
\usepackage[figuresright]{rotating}

\begin{document}


\title{The Chemical Separation of Eka-Hg from CERN W Targets in view of Recent
Relativistic Calculations\footnote{To be published in
Kerntechnik.}}


\author{D. Kolb}
\email{kolb@physik.uni-kassel.de} \altaffiliation{Fax:
+49-561-804-4006.}
 \affiliation{Department of
Physics, University GH Kassel, D-34109 Kassel, Germany}
\author{A. Marinov}
\affiliation{The Racah Institute of Physics, The Hebrew
University, Jerusalem 91904, Israel}
\author{G.W.A. Newton}
\affiliation{Heron's Reach, 382 Mossy Lea Road, Wrightington,
Lancashire, WN6 9RZ UK}
\author{R. Brandt}
\affiliation{Kernchemie, Philipps University, D-35041 Marburg,
Germany}


\date{October 11, 2004}

\begin{abstract}
In 1971  evidence for the production of element 112 via secondary
reactions in CERN W targets was obtained. The evidence was mainly
based on the observation of fission fragments in Hg sources
 separated from the W targets, on the measured masses of
the fissioning nuclei and on the assumption that element 112
(Eka-Hg) actually behaves like Hg in the chemical separation
process. This assumption is analyzed in view of recent
relativistic calculations of the electronic structure of element
112. It is shown that in the superheavy element region only the
chemistry of element 112 is similar to that of Hg.
\end{abstract}

\maketitle

\section{Introduction}
Back in 1971 long-lived fission fragments have been found in Hg
sources separated from two CERN W targets which were irradiated
with 24 GeV protons \cite{mar71a,mar71b}. Based on the
extrapolation of the periodic table the fission activity was
interpreted as due to production, via secondary reactions, of the
superheavy element with Z = 112. The masses of the fissioning
species were measured
 and  heavy masses like 272, 308 and 317-318 were
found and consistently interpreted as due to element 112 with
160-161 neutrons and various molecules of it \cite{mar84}. From
the measured mass of the produced superheavy nucleus cold fusion
reactions like $^{88}$Sr + $^{184}$W $\rightarrow$ $^{272}$112 and
$^{86}$Sr + $^{186}$W $\rightarrow$ $^{272}$112 were deduced
\cite{mar84}. The ordinary heavy ion reaction $^{88}$Sr +
$^{184}$W has been studied and characteristic X-rays of element
112 in coincidence with low energy particles (interpreted as
protons) and a very high-energy $\alpha$ particle (E$_\alpha$ =
12.16 MeV) in coincidence with a fission fragment were found
\cite{mar91a}. The results have been summarized in Refs.
\cite{mar91b,mar92,mar93}.

For many years it was difficult to understand these data
 and in particular the long lifetime of several weeks of the fission
activity, and the large deduced fusion cross sections of several
mb in the secondary reaction experiment, and in the nb range in
the ordinary heavy ion reaction. However, in recent years new
long-lived isomeric states with new radioactive decay properties
have been discovered \cite{mar96a,mar96b,mar01a,mar01b}. These
states are long-lived high spin super- and hyperdeformed isomeric
states.
 The evidence for the existence
of the long-lived high spin superdeformed isomeric states is based
in the first place   on the observation of relatively low energy
and about five orders of magnitude enhanced (as compared to
lifetime - energy relationship \cite{vio66}) $\alpha$ particles,
where the enhancement is consistent with the calculated
penetration through a barrier of a superdeformed nucleus, and the
alpha particles themselves were found to be in coincidence with
superdeformed band $\gamma$-ray transitions \cite{mar96a}.
Secondly,  very retarded proton activities were observed which
were interpreted as due to superdeformed to normal states
transitions \cite{mar96b}. The evidence for the existence of
similar hyperdeformed isomeric states is based on the observation
of high energy and 13 orders of magnitude retarded $\alpha$
particles in coincidence with superdeformed band $\gamma$ rays,
where the energy of the $\alpha$ particles fit with theoretical
predictions for hyperdeformed to superdeformed transition
\cite{mar01a}, and, in addition, on the observation of low
energies and about seven orders of magnitude enhanced $\alpha$
particles, where the low energies fit with theoretical predictions
for hyperdeformed to hyperdeformed transitions, and the large
enhancements fit with penetration calculations for such
transitions. The lifetimes of the observed high spin super- and
hyperdeformed isomeric states have been found to be much longer
than that of their corresponding normal ground states
\cite{mar01b}.

It was shown  that the existence of the super- and hyperdeformed
isomeric states enables one to consistently interpret the
production of the long-lived superheavy element with Z = 112
\cite{mar01b}. The long observed lifetime shows that  isomeric
state(s) rather than the normal ground state has been produced in
the reaction, and the large fusion cross section in the ordinary
heavy ion reaction is  due to the production of the compound
nucleus in
 a super- or hyperdeformed isomeric state. The shape of such a
state is similar to that of the projectile-target combination in
their touching point \cite{mar01b}. Therefore much less
overlapping and interpenetration are required in this case as
compared to the production of the compound nucleus in its normal
deformed, much more compact, shape. Secondly, in the secondary
reaction an additional effect is taking place
\cite{mar91b,mar92,mar93}. The projectile in this case is not a
nucleus in its ground state, but rather a fragment that has been
produced just within 2 x 10$^{-14}$ s before interacting with
another W nucleus in the target. During this short time it is
still at high excitation energy and quite deformed. Since the
Coulomb repulsion energy between the projectile and the target
nuclei, for tip to tip configuration, decreases as a function of
deformation, deformation has a large effect on the fusion cross
section as is well known from the sub-barrier fusion effect
\cite{stok80,iwa96}.

It is seen that a consistent interpretation to both the secondary
reaction experiment in CERN W targets and the ordinary $^{88}$Sr +
$^{184}$W reaction is possible, in terms of production of the
superheavy element with Z = 112 and N $\simeq$ 160 in an isomeric
state(s), provided that the chemical properties of element 112,
which were used in the chemical separation procedure, are similar
to those of Hg. Recently the question was raised if it is
justified to expect that element 112 will act like Hg, since some
relativistic calculations indicate that it might show properties
more like a noble gas \cite{per02}. (See also Refs.
\cite{pit75,set97} and references therein). The purpose of this
paper is to show that, including relativistic effects, only
element 112 is similar enough to Hg to follow it in the chemical
separation.

\section{The Chemical Separation Procedure that Isolated the Fission Activity
from the CERN W Targets}

In the original experiments \cite{mar71a,mar71b,mar84}  the above
fission fragments were found  in Hg sources that were separated
from two CERN W targets irradiated with 24-GeV protons, but not in
Au, Tl and Pb sources \cite{mar78}. The chemical procedure was
described in \cite{mar71a} and for the sake of completeness it is
summarized in Fig. 1. Fission fragments were seen several times
from the produced Hg sources \cite{mar71a,mar71b}.  The Hg source
from the W2 target was then electroplated, without applying any
voltage, on a small piece of Cu wire that was put in the ion
source of a mass separator and heated to about 300$^\circ$ C
\cite{fre77}. Based on experimental data and on extrapolation of
the periodic table this low temperature eliminates any element
with 90 $\leq$ Z $\leq$ 111. The masses of the fissioning species
were measured and interpreted as due to the atom and various
molecules of the $^{\sim272}$112 nucleus, where the molecules are
known to be common molecules formed with Hg itself \cite{mar84}.

    \begin{figure}[h]
\hspace*{-0.2cm}
\includegraphics[width=0.47\textwidth]{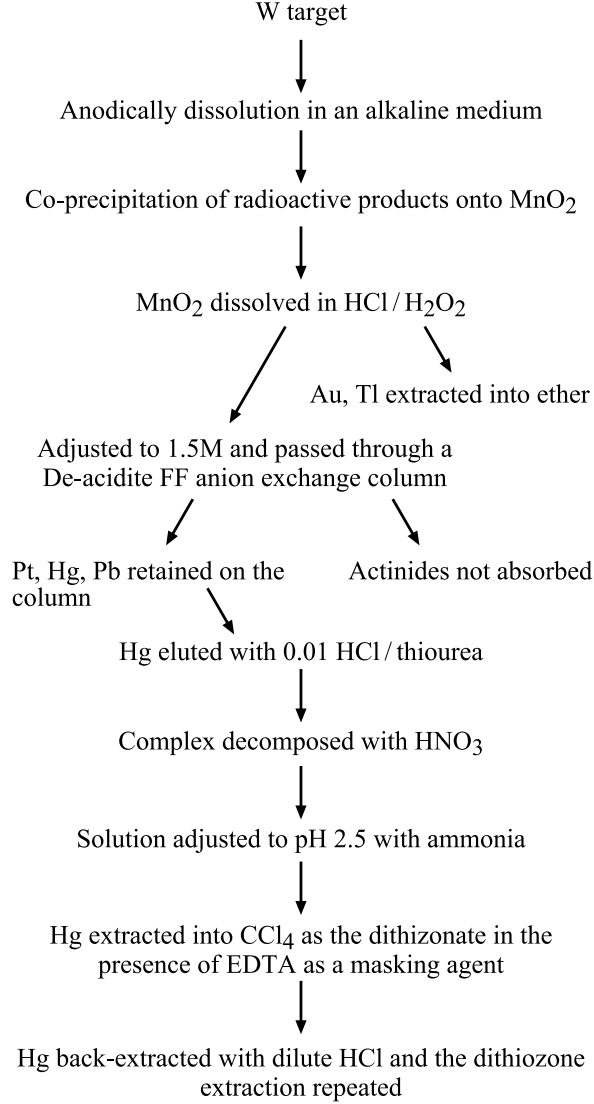}
 \caption{Block diagram of the chemical separation of Hg from the CERN W targets.}
\end{figure}

\section{The chemical separation procedure in view of recent
relativistic calculations}

 It is clear that the measured fission activity
basically followed the chemistry of Hg, otherwise one would have
readily lost about 500 atoms, which where responsible for the
observed fission fragments,  out of about 30 gm of W material,
 in the complex chemical
procedure. It must be due to a particular element. As mentioned
above, since in the periodic table element 112 is in the same
column as Hg having a closed s-d shell, it was concluded that the
fission fragments are due to this element. It should be mentioned
that this conclusion is also in accord with recent relativistic
calculations \cite{per02}. It is shown \cite{per02} that the
p$_{1/2}$ and s$_{1/2}$ level energy distance even increases from
Hg to element 112 and thus makes partial p$_{1/2}$ occupancy at
least as improbable as in Hg. Occupation of the p$_{1/2}$ shell by
 one electron or more (Z=113 and higher) or removal of at least one
electron from the s-d shell (Z below 112) would drastically change
the chemical and physical properties.  E.g. from the experience
with s-d atomic level structures in the periodic table we see that
the evaporation temperatures stay high as long as the s-d shell is
not closed, with a strong drop for the closed d$^{10}$s$^{2}$
configuration as in Hg and Z=112.  On the other hand the
ns$_{1/2}$ and (n-1)d$_{5/2}$ which were quite apart in Hg are now
almost energetically degenerate, because the 7s$_{1/2}$ orbital
becomes more bound due to a direct relativistic effect and the
6d$_{5/2}$ less bound due to indirect relativistic effects
(stronger shielding of the nuclear charge by the relativistically
enhanced deeper binding of the s$_{1/2}$ and p$_{1/2}$ orbitals).
The less bound 6d$_{5/2}$ makes it more reactive and the more
bound 7s$_{1/2}$ makes it more noble than in Hg. These two effects
may compensate and make chemical behaviors of Hg and element 112
similar. (For instance,  according to the relativistic
calculations \cite{per02} the binding energies (D$_{c}$) of Hg and
element 112 on Cu are almost the same). On the other hand the
chemistry of Hg cannot be similar to that of any other superheavy
element in this Z region. Therefore it may be concluded that the
chemical separation procedure described in Fig. 1, which was
followed by electroplating  on Cu (without applying any voltage)
and evaporation at low temperature of about 300$^\circ$ C, did
isolate element 112 and no other. It is essential to note that the
chemical procedure of Fig. 1 was done on Hg and element 112 at
various oxidation states, and not at an elemental state like in
\cite{yak03,sov04} (see also \cite{sar03,eic03}) where one is
basically sensitive to the volatility and the adsorption on Au
properties of the element. These properties might be more similar
to those of Rn than to Hg.

\section{Summary}

The chemical separation procedure that isolated Hg sources from
the CERN W targets is analyzed in view of recent relativistic
calculations for the atomic energy levels of element 112. It is
shown that the observed fission activity in the Hg sources with
its high measured masses of the fissioning species  can be
associated only with this element.

\section{Acknowledgements}

We appreciate very much the valuable discussions with J. L.Weil
and  N. Zeldes.

\end{document}